\documentclass[twocolumn,showpacs,aps,nofootinbib]{revtex4}

\begin{document}

 \newcommand{\bq}{\begin{equation}}
 \newcommand{\eq}{\end{equation}}
 \newcommand{\bqn}{\begin{eqnarray}}
 \newcommand{\eqn}{\end{eqnarray}}
 \newcommand{\nb}{\nonumber}
 \newcommand{\lb}{\label}
\newcommand{\PRL}{Phys. Rev. Lett.}
\newcommand{\PL}{Phys. Lett.}
\newcommand{\PR}{Phys. Rev.}
\newcommand{\CQG}{Class. Quantum Grav.}

\title{$f(R)$ term and geometric origin of the dark sector  in  Horava-Lifshitz gravity}

\author{Anzhong Wang}

\affiliation{
 GCAP-CASPER, Physics Department, Baylor University, Waco, TX 76798-7316, USA}

\date{\today}

\begin{abstract}

Inclusion of $f(R)$ term  in the action of  Horava-Lifshitz quantum gravity with projectability but 
without detailed balance condition is investigated, where $R$ denotes the 3-spatial dimensional 
Ricci scalar. Conditions for the spin-0 graviton to be free of ghosts and instability  are studied. 
The requirement that the theory reduce to general relativity in the IR makes the scalar mode 
unstable in the Minkowski background but stable in the de Sitter.  It is remarkable that the dark 
sector,  dark matter and dark energy, of the universe has a naturally geometric origin in such a 
setup. Bouncing universes can also be constructed. Scalar perturbations in the FRW backgrounds 
with non-zero curvature are presented.

\end{abstract}


\maketitle

\section{Introduction}
\renewcommand{\theequation}{1.\arabic{equation}} \setcounter{equation}{0}

There has been considerable interest recently on a  theory of quantum gravity proposed 
by Horava \cite{Horava}, motivated by Lifshitz scalar theory \cite{Lifshitz}. A free scalar field in 
$(d+1)$-dimensional flat spacetime is usually given by 
\bq
\lb{1.aa}
S_{\phi} =  \int{dt d^{d}x\left(\dot{\phi}^{2} + \phi\nabla^{2}\phi\right)},
\eq
which is Lorentz invariant,
\bq
\lb{1.a}
t \rightarrow \xi^{0}(t, {\bf x}), \; x^{i} \rightarrow \xi^{i}(t, {\bf x}), \; (i, j = 1, 2, 3).
\eq
The corresponding propagator is given by  $G_{\phi}(\omega, {\bf k}) =  {1/k^{2}}$,  indicating
 that the theory is not renormalizable. To improve the ultra-violet (UV) behavior, Lifshitz 
 introduced high-order spatial derivatives, $\phi \left(-\nabla^{2}\right)^{z}\phi$,
 into the action, and found that the resulted theory becomes    
 renormalizable
 for $z \ge d$ \cite{Visser}. An immediate consequence of these terms is that the theory is no longer
 Lorentz invariance, and $t$ and $x$  scale differently,
 \bq
\lb{1.1}
 t \rightarrow \ell^{z} t, \;\;\;  {x^{i}} \rightarrow \ell {x^{i}}.
\eq

Based on the above observations, Horava toke the point of view that Lorentz symmetry should appear as an 
emergent symmetry at long distances, but can be fundamentally 
absent at short ones \cite{Pav1,Pav2}. To realize such a  perspective,  Horava  started with
 the Arnowitt-Deser-Misner  (ADM) form of the metric,
 \bq
\lb{1.2}
ds^{2} = - N^{2}c^{2}dt^{2} + g_{ij}\left(dx^{i} + N^{i}dt\right)
     \left(dx^{j} + N^{j}dt\right),  
\eq
and imposed  the foliation-preserving  diffeomorphisms,  
\bq
\lb{1.3}
t \rightarrow f(t),\;\;\;  x^{i} \rightarrow \zeta^{i}(t, {\bf x}),
\eq
to be  denoted by Diff$_{{\cal{F}}}$(M). At low energies, the theory is expected to  flow to the IR fixed point
$z = 1$, whereby the Lorentz  invariance is ``accidentally restored."  

The effective speed of light in this  theory diverges in the UV regime, which could potentially resolve the 
horizon problem without invoking inflation \cite{KK}. The  spatial curvature is enhanced by higher-order 
curvature terms \cite{calcagni1,calcagni2,WM}, and this opens a new approach to the flatness problem and to 
a bouncing universe \cite{calcagni1,calcagni2,brand1,brand2}. 
In addition,  in the super-horizon region scale-invariant  
curvature perturbations can be produced without inflation \cite{Muk1,Muk2,Piao1,Piao2,Piao3,WMW}.   The perturbations 
become adiabatic during slow-roll inflation driven by a single scalar field, and the comoving curvature perturbation 
is constant  \cite{WMW}. Due to all  these remarkable features,  the HL theory has attracted lot of  attention 
lately, see, for example, Ref. \cite{HWW} and references therein.

To formulate his theory,   Horava started with  two conditions -- {\em detailed balance and projectability} 
\cite{Horava}. 
The detailed balance condition restricts the form of a general potential in a ($d+1$)-dimensional Lorentz action
to a specific form that can be expressed in terms of a d-dimensional action of a relativistic theory with Euclidean 
signature, whereby the number of independent-couplings is considerably limited. The projectability condition, on
the other hand,  
restricts the lapse function $N$
 to be space-independent, while the shift vector $N^{i}$ and 
the 3-dimensional metric $g_{ij}$ still depend on  
both time and space,
\bq
\lb{1.4}
N = N(t), \; N^{i} = N^{i}(t, x),\; g_{ij} = g_{ij}(t, x).
\eq
Clearly, this  
condition  is preserved by Diff$_{{\cal{F}}}$(M).  

It should be noted that, due to the restricted Diff$_{{\cal{F}}}$(M),  
one more degree of freedom appears
in the gravitational sector - a spin-0 graviton.  
In  particular, in the projectable case this mode is unstable in the Minkowski 
background  \cite{SVW1,SVW2,WM}.
As shown below, this is a generic feature of the theory with projectability condition.  
However,  this instability does not necessarily show up in physical environments  \cite{IM}.

In addition, it is also possible that   the legitimate  background in the HL theory is 
not Minkowski \footnote{Thanks to P. Horava for pointing it out to us.}. In particular,   recent observations show that our universe is currently de Sitter-like
\cite{obs}. Therefore, instead of the Minkowski, one may take the de Sitter  space as the background. 
As a matter of fact, it was shown recently that  the de Sitter  space is indeed stable in the framework of the HL theory
with projectability condition \cite{HWW,WWb}.

On the other hand, in the non-projectable case, this mode is also unstable \cite{BS}.  
However, if one includes   terms made of the spatial gradients of the lapse function, 
\bq
\lb{1.5}
a_{i} = \partial_{i}\ln N,
\eq
the mode can be rendered stable  \cite{BPS2}. However, an immediate price to pay for this is the inclusion of more than 
60 sixth-order derivative terms in the potential \cite{KP}. When matter, such as a scalar field, is included, the number
of such terms  increases dramatically. 
In addition, strong coupling may still exist  \cite{PS}, unless the scales
appearing in  higher order terms are much lower than the Planck scale \cite{BPS2}.   
 
It should be noted that strong couplings also happen in the version with projectability condition \cite{KA}. However,  Mukohyama recently 
showed that,
when nonlinear effects are taken into account, the spin-0 graviton decouples for spherically symmetric, static, vacuum 
spacetimes \cite{Mukc}. Together with Wu, we showed that this is also the case in the cosmological setting \cite{WWb}. 
As a  result, the relativistic continuation   $\lambda \rightarrow 1^{+}$  exists in the IR, whereby the 
strong coupling problem is circumvented.

The most general form  of the HL theory with projectability but without detailed balance condition was first developed
in    \cite{SVW1,SVW2} by Sotiriou, Visser and Weinfurtner (SVW), in which the highest order of spatial derivatives is  assumed
to be six, 
the minimal requirement for the theory to be power-counting renormalizable in (3+1)-dimensional spacetimes 
\cite{Horava,Visser}.  However, in principle there is nothing to prevent one to construct  actions with spatial derivatives 
higher than six. In 
addition, in condensed matter physics,  the critical exponent $z$ is also not necessarily an integer.
In this paper, we explore all these possibilities. Clearly, promoting all  terms to high orders is out of control.   
Instead, we consider high order derivatives only coming  from the Ricci scalar $R$, by simply replying  the  
third-order polynomial of $R$ in the SVW setup  by an  arbitrary function, $f(R)$, while keep the rest the same. 
As shown below, this simple generalization results in very rich physical phenomena, and in particular can give rise to dark matter
and dark energy. Since  in such a setup the origin of them is purely geometric, it automatically  explains
why these objects are ``dark".  Bouncing universes can be also easily constructed 
by properly choosing the form of $f(R)$. 


It is interesting to note that  in \cite{Mukb} it was advocated that the HL theory with projectability condition  has a  built-in dark matter 
component, due to the non-locality of the Hamiltonian constraint. In addition, $f({\cal{R}})$ models have
been  investigated in the framework of the HL theory \cite{Kluson1,Kluson2,Kluson3,Chai,Carl}, but in all these studies ${\cal{R}}$ 
is different from the 3D Ricci scalar $R$.  As a result, those $f({\cal{R}})$ models are fundamentally different from the ones   
studied here.  
For example, in \cite{Chai} ${\cal{R}}$ was token as
\bqn
\lb{1.6}
{\cal{R}} &=& K_{ij}K^{ij} - \lambda K^{2} + 2\mu \nabla_{\beta}\left(n^{\beta}\nabla_{\nu}n^{\nu}
- n^{\nu}\nabla_{\nu}n^{\beta}\right) \nb\\
&& - E^{ij}{\cal{G}}_{ijkl}E^{kl},
\eqn
where $n^{\nu}\; (\nu = 0, 1, 2, 3)$ is a unit vector perpendicular to the hypersurfaces of $t = $ Constant,  $\lambda$ and $\mu$
are two coupling constants,  ${\cal{G}}_{ijkl}$ is  the``generalized'' De Witt metric,  and $E^{ij}$ is given in terms of  a
super-potential. For detail, we refer readers to \cite{Kluson1,Kluson2,Kluson3,Chai,Carl}. Clearly, for any choice of $\lambda,\; \mu$ and $E^{ij}$,
due to the presence of the first term on the right-hand side of Eq.(\ref{1.6}), ${\cal{R}}$ cannot reduce to the
three-spatial dimensional  Ricci scalar $R$. In addition, due to its presence, the corresponding theory usually involves high order time derivatives
\cite{Kluson1,Kluson2,Kluson3,Chai,Carl}, a situation that was avoided in the first place by Horava,  in order to circumvent  the ghost problem \cite{Horava}. 

 The rest of the paper is organized as follows: In Section II, we present the Hamiltonian and momentum constraints, the dynamical equations,
 and the conservation laws of energy and momentum, after including  the $f(R)$ term in the SVW setup. To separate the effect of this term
 from others introduced in \cite{SVW1,SVW2}, in this section we  only consider the $f(R)$ term. In Section III, we study the  
 stability of the spin-0 graviton  in the Minkowski background, and obtain the stability condition.   We show explicitly that it is the requirement
 that the theory reduce to its relativistic limit at the IR leads to the instability of  the spin-0 graviton.  This is also true when all the other terms
 are included. In Section IV, we investigate its applications to cosmology, by first writing the corresponding Friedmann equations of the FRW 
 universe with arbitrary curvature, and then their linear  scalar perturbations.  From these expressions, it can be easily shown that the de Sitter
 spacetime is stable, similar to the case without the $f(R)$ term \cite{HWW}. In this section, we also generalize our studies to include all the other
 terms introduced in \cite{SVW1,SVW2}. In Section V, we present our main results and give some 
 concluding  remarks.

\section{Inclusion of the $f(R)$ Term}
\renewcommand{\theequation}{2.\arabic{equation}} \setcounter{equation}{0}

To see the role that the term $f(R)$ may play, we shall first neglect all the other terms
constructed in  \cite{SVW1,SVW2}, and simply consider the action, 
 \bqn \lb{2.4}
S = \zeta^2\int dt d^{3}x N \sqrt{g} \left({\cal{L}}_{K}  + f(R)
+\zeta^{-2} {\cal{L}}_{M} \right),
 \eqn
where $g={\rm det}\,g_{ij}$,
 \bq \lb{2.5}
{\cal{L}}_{K} = K_{ij}K^{ij} - \lambda  K^{2}, 
 \eq
$\zeta^{2} = 1/{16\pi G}$, $\lambda$ is a dynamical coupling constant, and $f(R)$ is an arbitrary function of 
$R$. 
However, to have the theory power-counting renormalizable, it is necessary  to include terms that are equal or
higher than $R^{3}$.  The  extrinsic curvature $K_{ij}$ is defined as
 \bq \lb{2.6}
K_{ij} = \frac{1}{2N}
\left(- \dot{g}_{ij} + \nabla_{i}N_{j} +
\nabla_{j}N_{i}\right),
 \eq
where $N_{i} = g_{ij}N^{j}$, and the covariant derivatives  refer to the three-metric $g_{ij}$.

Variation with respect to the lapse function $N(t)$ yields the global
Hamiltonian constraint,
 \bq \lb{eq1}
\int{ d^{3}x\sqrt{g}\left({\cal{L}}_{K} - f(R)\right)}
= 8\pi G \int d^{3}x {\sqrt{g}\, J^{t}},
 \eq
where
 \bq \lb{eq1a}
J^{t} \equiv 2\frac{\delta({N\cal{L}}_{M})}{\delta N}.
\eq
 Variation with respect to   $N^{i}$ yields the
super-momentum constraint,
 \bq \lb{eq2}
\nabla_{j}\pi^{ij} = 8\pi G J^{i},
 \eq
where 
$\pi^{ij} $ and 
$J^{i}$
are defined as
 \bqn \lb{eq2a}
\pi^{ij} &\equiv& \frac{\delta{(N\cal{L}}_{K})}{\delta\dot{g}_{ij}} = 
- K^{ij} + \lambda K g^{ij}, \; \nb\\
J^{i} &\equiv& - N\frac{\delta{\cal{L}}_{M}}{\delta N_{i}}.
\eqn
Varying with respect to $g_{ij}$, on the other hand, leads to the
dynamical equations,
 \bqn \lb{eq3}
&&
\frac{1}{N\sqrt{g}}\left(\sqrt{g}\pi^{ij}\right)^{\displaystyle{\cdot}}
= 2 \lambda K K^{ij} -2\left(K^{2}\right)^{ij} + \frac{1}{2} {\cal{L}}_{K}g^{ij}    + F^{ij}
\nb\\
& &  ~~~~ + \frac{1}{N}\nabla_{k}\left(N^k \pi^{ij} - \pi^{ki}N^{j} - \pi^{kj}N^{i}\right)
+ 8\pi G \tau^{ij},  ~~~ 
 \eqn
where $\left(K^{2}\right)^{ij} \equiv K^{il}K_{l}^{j}$,
and
\bqn
\lb{eq3a} 
\tau^{ij} & \equiv & \frac{2}{\sqrt{g}}\frac{\delta \left(\sqrt{g}
 {\cal{L}}_{M}\right)}{\delta{g}_{ij}},\nb\\
 F^{ij} &\equiv&  \nabla^{i}\nabla^{j}F - g^{ij}\nabla^{2}F - F R^{ij} +  \frac{1}{2}fg^{ij}, 
\eqn
with $F \equiv df(R)/dR$.

The matter quantities $(J^{t}, \; J^{i},\; \tau^{ij})$ satisfy the
conservation laws,
 \bqn \lb{eq4a} & &
 \int d^{3}x \sqrt{g} { \left[ \dot{g}_{kl}\tau^{kl} -
 \frac{1}{\sqrt{g}}\left(\sqrt{g}J^{t}\right)^{\displaystyle{\cdot}}
  \right. }\nb\\
 & &  \left.  \;\;\;\;\;\;\;\;\;\;\;\;\;\;\;\;\;\; +  \frac{2N_{k}}
 {N\sqrt{g}}\left(\sqrt{g}J^{k}\right)^{\displaystyle{\cdot}}
 \right] = 0,\\
\lb{eq4b} & & \nabla^{k}\tau_{ik} -
\frac{1}{N\sqrt{g}}\left(\sqrt{g}J_{i}
\right)^{\displaystyle{\cdot}} - \frac{N_{i}}{N}\nabla_{k}J^{k} \nb\\
& & \;\;\;\;\;\;\;\;\;\;\; - \frac{J^{k}}{N}\left(\nabla_{k}N_{i}
- \nabla_{i}N_{k}\right) =
 0.
\eqn

\section{Spin-0 Graviton in Minkowski Background}
\renewcommand{\theequation}{2.\arabic{equation}} \setcounter{equation}{0}

It can be shown that Minkowski spacetime $ds^{2} = - dt^{2} + \delta_{ij}dx^{i}dx^{j}$ is a solution
of the vacuum field equations, provided that  $ f(0) = 0$. 
Considering its linear scalar perturbations,  
 \bq
  \lb{3.1}
\delta{g}_{ij} =   - 2\psi\delta_{ij} + 2E_{,ij}, \; 
\delta{N}_{i} =    B_{,i} ,   \;  \delta{N} =  \phi, 
\eq
 in  the quasi-longitudinal gauge \cite{WM},
$\phi = 0 = E$,
we find that 
to second order the   action without matter takes the form,
\bqn
\lb{3.3}
S^{(2)}_{g} &=&  \zeta^{2}\int{d\eta d^{3}x N\sqrt{g}\Big({\cal{L}}^{(2)}_{K} + f^{(2)}(R)\Big)}\nb\\
&=& \zeta^{2}\int{dt d^{3}x  \Big[\big(1-3\lambda\big)\Big(3\dot{\psi}^{2}   + 2 \dot{\psi}\partial^{2}B}\Big)
\nb\\
& &  + (1-\lambda)B\partial^{4}B + 2\gamma \psi\partial^{2}\psi - {8\omega \zeta^{-2}}  \psi\partial^{4}\psi\Big], ~~~~~~
\eqn
 where 
 \bq
 \lb{3.3a}
 \gamma \equiv - f'(0),\;\;\;
 \omega \equiv - \zeta^{2} f''(0). 
 \eq
 Variations with respect to $B$ and $\psi$ yield,
 respectively, 
 \bqn
 \lb{3.4a}
& &  (1-3\lambda)\dot{\psi}_{k} - (1-\lambda) k^{2}B_{k} = 0,\\
 \lb{3.4b}
 & & \big(3\lambda -1\big)\left(3\ddot{\psi}_{k} - k^{2}\dot{B}_{k}\right) 
 = 2k^{2}\left(\gamma + 4\omega k^{2}/ \zeta^{2}\right)\psi_{k}, ~~~~~
 \eqn
 in the momentum space. When $\lambda \not= 1$, from Eq. (\ref{3.4a}) 
 we can express $B_{k}$ in terms of $\psi_{k}$, and
then   Eq. (\ref{3.3}) becomes 
 \bq
 \lb{3.5}
 S^{(2)}_{g} = \frac{2\zeta^{2}}{c^{2}_{\psi}}\int{dt d^{3}x \Big(\dot{\psi}_{k}^{2} - \omega^{2}_{\psi}\psi^{2}_{k}\Big)},
 \eq
 with 
 \bqn
 \lb{3.5a}
 c^{2}_{\psi} &\equiv& \frac{\lambda -1}{3\lambda - 1},\nb\\
 \omega^{2}_{\psi} &\equiv&  c^{2}_{\psi}k^{2}\left(\gamma + \frac{4\omega k^{2}}{\zeta^{2}}\right).
 \eqn
Hence, $\psi_{k}$ satisfies 
 $\ddot{\psi}_{k} + \omega^{2}_{\psi} \psi_{k} = 0$, which has stable solutions only when $\omega^{2}_{\psi} > 0$. Clearly, this is true
in the IR limit  only when 
\bq
\lb{3.5ab}
\lambda > 1,\;\;\;
\gamma > 0. 
\eq
These conditions also assure the kinetic term in 
$S^{(2)}_{g}$ always positive, that is, 
 free of ghosts.  When $\lambda  =1$, from Eq.(\ref{3.4a}) we find that ${\psi}_{k}$ and $\dot{B}_{k} $ are independent of $t$, and  so are 
  the two gauge-invariant quantities \cite{WM} $\Phi_{k} = \dot{B}_{k}$ and $\Psi_{k} = \psi_{k}$. Therefore,  the spin-0
  graviton is stable when $\lambda  = 1$. 
  
  Note that the addition of other terms presented in \cite{SVW1,SVW2} does not change the IR
  behavior, and their contributions only change the expression $\omega^{2}_{\psi}$ to the form,
  \bq
  \lb{3.5b}
  \omega^{2}_{\psi} = c^{2}_{\psi}k^{2}\left(\gamma + \frac{\left(4\omega - 3g_{3}\right) k^{2}}{\zeta^{2}}
  + \frac{\left(3g_{8} - 8g_{7}\right)k^{4}}{\zeta^{4}}\right),
  \eq
 where $g_{i}$ are the coupling constants introduced in \cite{SVW1,SVW2}.
 In addition, the condition that the theory reduces to general relativity (GR) requires $\gamma = -1$, 
 a condition that was assumed in  \cite{SVW1,SVW2}. This explains why  in the SVW setup    the spin-0 mode is not stable \cite{WM}.
 Our analysis given above shows that the instability is a generic feature of the HL theory with projectability condition. However, this instability 
 does not necessarily show up in a physical environment, as longer as some conditions are satisfied \cite{IM}.  In particular, 
 if $L/|c_{\psi}| > t_{J}$, the instability will
not show up, where $L$ is the  length scale of interest, and  $t_{J}$ denotes the timescale of   Jeans instability \cite{IM}. 
In addition, the linear instability is stabilized by higher derivative
terms if $|c_{\psi}| < 1/(LM_{*})$, where $M_{*}$ is the energy scale suppressing higher derivative
terms. In the current setup, it is  given by $M_{*} \simeq |g_{3}|^{-1/2}M_{pl}, \;  |g_{s\ge 5}|^{-1/4}M_{pl}$.
The linear instability can be also  tamed by Hubble friction, if
$L/|C_{\psi}|  > 1/H$,  where $H$ is the Hubble expansion rate at the time of interest. 
Therefore, even a solution is not stable, one may properly choose the form of the function $f(R)$, so that at least one of 
these conditions is satisfied.  Then, the corresponding instability will not show up. 

It is also interesting to note that, although the SVW setup is not stable in the Minkowski background, it is stable  
in the de Sitter  \cite{HWW}. This is also true in the current setup. Then, one may consider the instability found
in the Minkowski background is just an indication that the latter is not the ground state of the theory.
Instead, the  legitimate  background in the HL theory might be the de Sitter spacetime. This point of view is further
supported by  current observations  that our universe is currently de Sitter-like \cite{obs}.

 \section{Cosmological models}
\renewcommand{\theequation}{4.\arabic{equation}} \setcounter{equation}{0}

\subsection{FRW Background} 

The homogeneous and isotropic universe is described by the FRW
metric, $ ds^{2} = - dt^{2} + a^{2}(t)\gamma_{ij}dx^{i}dx^{j}$
where $\gamma_{ij}={\left(1 + \frac{1}{4}\kappa r^{2}\right)^{-2}}
{\delta_{ij}}$, with $\kappa = 0, \pm 1$. For this metric, $\bar K_{ij}
= - a^{2}H \gamma_{ij}$ and $\bar R_{ij} = 2\kappa\gamma_{ij}$, where
$H = \dot{a}/a$ and an overbar denotes a background quantity. 
Setting \cite{WM}
 \bq\label{mq}
\bar{J}^t=-2\bar\rho,~~ \bar{J}^i=0,~~ \bar \tau_{ij} = \bar p\,
\bar g_{ij},
\eq
where $\bar\rho$ and $\bar p$ are the total density and pressure, we find that 
 the Hamiltonian constraint (\ref{eq1}) reduces to 
  \bq \lb{4.1a}
 \tilde{\lambda} H^{2}   = 
\frac{8\pi G}{3}\bar \rho   - \frac{1}{6} f(\bar{R}),  
 \eq
 where $\bar{R} = 6\kappa/a^{2}$,  and $\tilde{\lambda} \equiv \left(3\lambda - 1\right)/2$.
  It can be shown that the momentum constraint (\ref{eq2}) is 
 satisfied identically, while   the dynamical equations (\ref{eq3})  yield,
   \bq \lb{4.1b}
 \tilde{\lambda} \frac{\ddot{a}}{a} =   - {4\pi G\over3}(\bar\rho+3\bar p)
-  \frac{1}{6} f(\bar{R}) + \frac{\kappa}{a^{2}} F(\bar{R}).  
 \eq
 The conservation of momentum (\ref{eq4b}), on the other hand, is also satisfied identically,
 while the conservation of energy (\ref{eq4a}) gives
  \bq \lb{4.1c}
\dot{\bar{\rho}} + 3H \left(\bar\rho +\bar p \right) = 0,
 \eq
which can be obtained from Eqs. (\ref{4.1a}) and (\ref{4.1b}).
When $\kappa = 0$ we have $\bar{R} = 0$, and these two equations reduce exactly to these
 given in GR with  a cosmological constant $\Lambda \equiv - f(0)/2$. 
In addition,   Eqs. (\ref{4.1a}) and (\ref{4.1b}) also admit a de Sitter  point that corresponds 
to a vacuum solution ($\bar{\rho} = \bar{p} = 0$), at which
$ f({R}) - {R}f'({R}) = -2\Lambda$,
 which has the solution $f({R}) = -2\Lambda - \gamma {R}$. 
 
 In the rest of  this paper,  we shall  consider only  the cases  where $ \kappa \not= 0$. Then, Eqs. (\ref{4.1a}) and (\ref{4.1b})
are quite different from the ones of 4-dimensional $f(R^{(4)})$ models \cite{4fR1,4fR2}. In particular, 
since $R^{(4)}$ contains second derivatives of $a$, the generalized Friedmann equations
are usually high-order differential equations in the 4-dimensional $f(R^{(4)})$ models \cite{4fR1,4fR2}. But, 
in the HL theory   $\bar{R} [= 6\kappa/a^{2}]$ is a function of  $a$ only. As a result, $f(\bar{R})$ is  a   polynomial of $a$ only, too. 
Therefore, in the current setup these 
terms act  like sources. For example, if $f({R}) \propto   {R}^{3/2}$, then from  Eq. (\ref{4.1a})  we 
can see that this corresponds to  dark matter. Note that to have $F({R})$ real, in this case we must assume $\kappa > 0$.  
 In addition, 
 for $f({R})  \propto   {R}^{n}$  Eq. (\ref{4.1a}) shows that $H^{2} \propto a^{-2n}$. Then, with $n < 1$ this term
 can mimic dark energy. On the other hand, for $n \ge 2$ it can give rise to a bouncing universe 
 \cite{calcagni1,calcagni2,brand1,brand2}.  Since
all of these terms are purely geometric, they do not subjected to the energy conditions
\cite{HE72} and matter instabilities. 
Note that the unification of the dark sector  in  $f(^{(4)}R)$  models was first considered in \cite{NS03}.

\subsection{Linear Scalar Perturbations}
 
Consider scalar perturbations in the quasi-longitudinal gauge
\cite{WM}, 
\bq
\lb{equ0}
\delta N = 0,\; \delta N_{i} = a^{2}B_{|i},\; \delta g_{ij}  = -2a^{2}\psi\gamma_{ij}, 
\eq
it can be shown that
  the Hamiltonian constraint (\ref{eq1}) to first order yields,
\bqn
\lb{eq1aA}
& & \int{\sqrt{\gamma} d^{3}x\Big[F(\bar{R})\left(\vec{\nabla}^{2}\psi + 3\kappa \psi\right)
- \tilde{\lambda}{\cal{H}}\left(\vec{\nabla}^{2}B + 3\psi'\right)} \nb\\
& & ~~~~~~~~~~~~~~~ = 4\pi G a^{2} \int{\sqrt{\gamma} d^{3}x \delta\mu},
\eqn
where $\delta\mu = -\delta J^{t}/2$, $\vec{\nabla}^{2}f \equiv \gamma^{ij} f_{|ij}$, and $|_{i}$ 
denotes the covariant derivative with respect to $\gamma_{ij}$. The momentum constraint (\ref{eq2})
takes the form,
\bq
\lb{eq2aA}
\left(3\lambda - 1\right)\psi' - 2\kappa B + \left(\lambda - 1\right)\vec{\nabla}^{2}B
= 8\pi G a q,
\eq
with $\delta J^{i} = a^{-2}q^{|i}$. The dynamical equations (\ref{eq3}), on the other hand, yield,
\bqn
\lb{eq3aa}
& & \psi'' + 2{\cal{H}}\psi' - \frac{2F(\bar{R})}{3(3\lambda - 1)}\left(\vec{\nabla}^{2}\psi + 3\kappa \psi\right)\nb\\
& &  ~~~~~~~~~~ +  \frac{8F'(\bar{R})}{3(3\lambda - 1)a^{2}}\left(\vec{\nabla}^{4}\psi - 3\kappa\vec{\nabla}^{2}\psi 
+ 9 \kappa^{2}\psi\right) \nb\\
& &  ~~~~~~~~~~ + \frac{1}{3} \vec{\nabla}^{2}\left(B' + 2{\cal{H}}B\right)  = \frac{8\pi G a^{2}}{3\lambda - 1}\delta{\cal{P}},\\
\lb{eq3ab}
& & B' + 2{\cal{H}}B =  F(\bar{R})\psi  - \frac{4F'(\bar{R})}{a^{2}}\left(\vec{\nabla}^{2}\psi + 3\kappa \psi \right) \nb\\
& & ~~~~~~~~~~~~~~~~~~ - 8\pi G a^{2}\Pi,
\eqn
where 
\bqn
\lb{eq3ac}
\delta\tau^{ij} &=&  a^{-2} \left[\left(\delta{\cal{P}} + 2\bar{p}\psi\right)\gamma^{ij} + \Pi^{|<ij>}\right], \nb\\
\Pi^{|<ij>} &=&  \Pi^{|ij} -  \frac{1}{3}\gamma^{ij} \vec{\nabla}^{2}\Pi. 
\eqn
The conservation laws give,
 \bqn 
 \lb{eq4aa}
& & \int \sqrt{\gamma} d^{3}x \Big[\delta\mu' + 3{\cal H}
\left(\delta{\cal P} + \delta\mu\right)
-3 \left(\bar\rho + \bar p\right){\psi}' \Big] =
0,\nb\\ 
& & ~~~~~~~~~~~~~~~~~~~~~~~~~~~~~~~~~~~~~~~~~~~~~~~~~~~~~~\\ 
\lb{eq4bb}
& & q'+3 {\cal H}q  - a\delta{\cal P} -
{2a\over3}\left( \vec\nabla ^{2}+3k \right)\Pi = 0. 
 \eqn
 Following \cite{HWW} it can be shown from Eq.(\ref{eq3aa}) that the de Sitter background is indeed  stable
  in the current setup.

\subsection{Inclusion of Other Terms Upto Six-Order}

As mentioned above, one can add  the   terms \cite{SVW1,SVW2}   
\bqn
\lb{eq0}
\delta{\cal{L}}_{{V}} &=&   \frac{\gamma_{1}}{\zeta^{2}}R_{ij}R^{ij}  
+  \frac{1}{\zeta^{4}} 
\left(\gamma_{2}  R\;
R_{ij}R^{ij}
+   \gamma_{3}  R^{i}_{j} R^{j}_{k} R^{k}_{i} \right)\nb\\
& & 
 + \frac{1}{\zeta^{4}} 
\left [\gamma_{4}R\nabla^{2}R +  \gamma_{5}
\left(\nabla_{i}R_{jk}\right)\left(\nabla^{i}R^{jk}\right)\right],
\eqn
into  
(\ref{2.4}), so that up to the sixth-order derivatives the  action is the most general one,  where   
$\gamma_{s}$'s are the coupling constants, defined as 
$\left\{\gamma_{s}\right\} \equiv \left(g_{3}, g_{5}, g_{6}, g_{7}, g_{8}\right)$. 
Then,  the  Hamiltonian 
constraint can be obtained from Eq.(\ref{eq1}) with the replacement of $f(R)$ by $ - {\cal{L}}_{{V}} $, where
 \bq
 \lb{4.3aa}
 {\cal{L}}_{{V}} \equiv \delta{\cal{L}}_{{V}} - f(R), 
 \eq
 while the super-momentum constraint (\ref{eq2}) and the conservation
 laws, (\ref{eq4a}) and (\ref{eq4b}) 
  remain the same. The dynamical equations are also
 given by Eq.(\ref{eq3}), but now with 
 \bq
 \lb{4.3}
 F^{ij} =  F^{ij}_{f(R)} + \sum^{5}_{s=1}{{\gamma_{s} }{\zeta^{-2m_{s}}} }\,
\left({\cal{F}}_{s}\right)^{ij},
\eq
 where $F^{ij}_{f(R)}$ is   defined by Eq. (\ref{eq3a}),  $m_{s} =( 1, 2, 2, 2,
2)$, 
and $ \left({\cal{F}}_{s}\right)^{ij} = \left(F_{3}, F_{5}, F_{6}, F_{7}, F_{8}\right)^{ij}$,
where $ \left({{F}}_{s}\right)^{ij}$'s are defined in \cite{SVW1,SVW2,WM}.

Due to the inclusion of   the above terms in the action, 
Eqs. (\ref{4.1a}) and (\ref{4.1b}) become
  \bqn
   \lb{4.1aa}
& & \tilde{\lambda} H^{2}   = 
\frac{8\pi G}{3}\bar \rho   - \frac{1}{6} f(\bar{R}) + \frac {2\beta_{1}\kappa^{2}}{a^{4}} 
+  \frac{4\beta_{2}\kappa^{3}}{a^{6}},\\
   \lb{4.1bb}
& & \tilde{\lambda} \frac{\ddot{a}}{a} =   - {4\pi G\over3}(\bar\rho+3\bar p)
-  \frac{1}{6} f(\bar{R}) + \frac{\kappa}{a^{2}} F(\bar{R})  \nb\\
& & ~~~~~~~~ - \frac{2\beta_{1}\kappa^{2}}{a^{4}}  -  \frac{8\beta_{2}\kappa^{3}}{a^{6}},
 \eqn
where 
\bq
\lb{4.1cc}
\beta_{1} \equiv \frac{\gamma_{1}}{\zeta^{2}},\;\;\;
 \beta_{2} \equiv \frac{3\gamma_{2} + \gamma_{3}}{\zeta^{4}}.
 \eq
Similarly,  one can obtain  the corresponding scalar perturbations from the above and \cite{WM}, which will not
be presented here. 
  
  \section{Conclusions}
\renewcommand{\theequation}{5.\arabic{equation}} \setcounter{equation}{0}

We studied the efforts of spatial derivatives higher than six, and the possibility 
of   the critical exponents $z$ being non-integer, in 
the  HL gravity with projectability 
but without detailed balance condition, by replying  the third-order polynomial of the Ricci scalar in the SVW setup  \cite{SVW1,SVW2}
by an  arbitrary function, $f(R)$, while keeping the rest the same. The requirement that the theory reduce 
 to GR in the IR makes scalar perturbations unstable in the Minkowski background, but stable in the de Sitter. 
 This might imply that the legitimate  background in the HL theory is 
not Minkowski spacetime, but the de Sitter.

This simple generalization leads to 
very rich physical phenomena. In particular, it naturally gives rise to dark matter
and dark energy. 
 We also studied scalar perturbations in the FRW backgrounds with non-zero curvature, and presented the general formulas, which will
 make the studies of perturbations in such a setup considerably  easier.   
It would be very interesting to study  their applications to large-scale 
structure formation,  early universe, and investigate constraints of the models from solar system tests.

When high-order curvature terms, $f(^{(4)}R)$, are included, it was shown that spin-2 massive models also exist and are ghosts \cite{BCDN}.
In the SVW setup of the HL theory, we have shown recently that vector perturbations vanish, while the only remaining tensor modes are the 
massless spin-2 modes \cite{Wang}.  Therefore, it would be very interesting to see the effects of the $f(R)$ terms on the vector and tensor
modes in the current setup. 
 
 In addition, dark matter and dark energy have been studied extensively \cite{DM,DE1,DE2}, and current observational data impose serve constraints
on various models \cite{obs}.  It is also very interesting to fit the data to our current models by using the MCMC code, recently   developed by us
 \cite{GWW}, including  the studies of the growth factor of perturbations \cite{GIW}. 
  
 Strong couplings \cite{KA} may exist here, too. One way is to provoke the Vainshtein mechanism \cite{Vain}, similar to what was done
in  \cite{Mukc,WWb}.

~\\{\bf Acknowledgements:} The author would like to thank    P. Horava, R. Maartens, A. Papazoglou, 
and D. Wands for valuable suggestions and comments. Part of this work was done when the author attended
the advanced workshop ``Dark Energy and Fundamental Theory," supported by the Special Fund for Theoretical Physics
 from the National Natural Science Foundation of China by the grant 10947203. The author would like to express his gratitude
 to Jianxin Lu for his organization of the Conference and hospitality.  
 This work was supported in part by DOE Grant, DE-FG02-10ER41692.


\end{document}